\documentclass[fleqn,usenatbib]{mnras}

\usepackage{newtxtext,newtxmath}
\usepackage[T1]{fontenc}

\DeclareRobustCommand{\VAN}[3]{#2}
\let\VANthebibliography\thebibliography
\def\thebibliography{\DeclareRobustCommand{\VAN}[3]{##3}\VANthebibliography}

\usepackage{graphicx}	% Including figure files
\usepackage{amsmath}	% Advanced maths commands
\usepackage{amssymb}	% Extra maths symbols

\usepackage{booktabs}
\usepackage{natbib}
\bibpunct{(}{)}{;}{a}{}{,}
\usepackage{supertabular}
\usepackage{lscape}
\usepackage{array}
\usepackage{float}
\usepackage{varioref}
\usepackage{multirow}
\usepackage{epstopdf}
\usepackage{comment}
\usepackage{xfrac}
\usepackage{rotating}
\usepackage{threeparttable, tablefootnote}
\usepackage{etoolbox}
\usepackage{siunitx}
\usepackage{tabularx}
\title[Extending the Breakthrough Listen star survey]{Extending the Breakthrough Listen nearby star survey to other stellar objects in the field}

\author[B. S. Wlodarczyk-Sroka et al.]{
Bart S. Wlodarczyk-Sroka,$^{1}$\thanks{E-mail: bart.wlodarczyk-sroka@postgrad.manchester.ac.uk}
M. A. Garrett$^{1,2}$
and A. P. V. Siemion$^{3,1}$
\\
$^{1}$Jodrell Bank Centre for Astrophysics, Department of Physics \& Astronomy, Alan Turing Building, The University of Manchester, M13 9PL, United Kingdom\\
$^{2}$Leiden Observatory, Leiden University, PO Box 9513, 2300 RA Leiden, The Netherlands\\
$^{3}$Department of Astronomy, University of California Berkeley, Berkeley CA 94720, USA
}

\date{Accepted XXX. Received YYY; in original form ZZZ}

\pubyear{2020}

\begin{document}
\label{firstpage}
\pagerange{\pageref{firstpage}--\pageref{lastpage}}
\maketitle

\begin{abstract}
We extend the source sample recently observed by the Breakthrough Listen Initiative by including additional stars (with parallaxes measured by {\textit{Gaia}}) that also reside within the FWHM of the GBT and Parkes radio telescope target fields. These stars have estimated distances as listed in the extensions of the {\textit{Gaia}} DR2 catalogue. Enlarging the sample from 1327 to 288315 stellar objects permits us to achieve substantially better Continuous Waveform Transmitter Rate Figures of Merit (CWTFM) than any previous analysis, and allows us to place the tightest limits yet on the prevalence of nearby high-duty-cycle extraterrestrial transmitters. The results suggest $\lesssim 0.0660${\raisebox{0.5ex}{\tiny$^{+0.0004}_{-0.0003}$}}\% of stellar systems within 50~pc host such transmitters (assuming an EIRP $ \gtrsim 10^{13}$~W) and $\lesssim 0.039${\raisebox{0.5ex}{\tiny$^{+0.004}_{-0.008}$}}\% within 200~pc (assuming an EIRP $\gtrsim 2.5 \times 10^{14}$~W). We further extend our analysis to much greater distances, though we caution that the detection of narrow-band signals beyond a few hundred pc may be affected by interstellar scintillation. The extended sample also permits us to place new constraints on the prevalence of extraterrestrial transmitters by stellar type and spectral class. Our results suggest targeted analyses of SETI radio data can benefit from taking into account the fact that in addition to the target at the field centre, many other cosmic objects reside within the primary beam response of a parabolic radio telescope. These include foreground and background galactic stars, but also extragalactic systems.  With distances measured by {\textit{Gaia}}, these additional sources can be used to place improved limits on the prevalence of extraterrestrial transmitters, and extend the analysis to a wide range of cosmic objects.
\end{abstract}

\begin{keywords}
Astronomical instrumentation, methods and techniques -- Astrometry -- Stars: general -- radio continuum: general
\end{keywords}

%%%%%%%%%%%%%%%%%%%%%%%%%%%%%%%%%%%%%%%%%%%%%%%%%%

%%%%%%%%%%%%%%%%% BODY OF PAPER %%%%%%%%%%%%%%%%%%

\section{Introduction} \label{1}

The Breakthrough Listen Initiative \citep{Worden2017} has recently conducted a state-of-the-art SETI (Search for Extraterrestrial Intelligence) survey of nearby stars, all of which are located within 50~pc of the Earth \citep{enriquez2017, Price2020}. The aim is to detect or place limits on the incidence of artificial transmitters of extraterrestrial origin. More specifically, the search focuses on the detection of very narrow-band radio signals ($\lesssim \rm{few~Hz}$ in width) which are not known to be generated by any natural, astrophysical processes.  

The initial target list of nearby stars for Breakthrough Listen was defined by \cite{isaacson2017}, comprising a broad sample of 1709 nearby main-sequence stars. The sample was selected by including all stars located within a distance of 5~pc contained within the Gliese and RECONS catalogues, and aiming to achieve a broad sampling of stars from the Hipparcos catalogue at distances between 5 and 50~pc. This was achieved by defining domains in B-V colour and V-band magnitude, along the entire length of the main sequence and selecting the 100 nearest stars from each domain. In addition to the main sequence targets, a further 100 stars in the 5-50~pc range were selected from another domain constructed to cover giant and sub-giant stars. The focus on nearby stars in the \citet{isaacson2017} sample allowed relatively low limits to be placed on the power of any transmitters, in accordance with the inverse square law. 

A sample of 692 of these stars was first observed by \cite{enriquez2017} with the Green Bank 100-m Telescope (hereafter GBT) using the L-band receiver (1.1-1.9~GHz). Each target star was observed for 3 scans each of 300 seconds in length, interspersed with 300 second scans of a variety of off-source secondary fields. The latter were chosen to lie well beyond the primary beam and side lobe response of the GBT in the target fields. This "ON-OFF" observing strategy serves to reduce the number of false positives (e.g. terrestrial radio frequency interference) - in particular, an extraterrestrial signal associated with a nearby star should only appear in the data associated with the three target (ON) scans but not in the offset (OFF) scans. Due to the high prevalence of RFI (Radio Frequency Interference) encountered in the data, even while employing the ON-OFF observing strategy, candidate "events" were limited in the first instance to a signal-to-noise ratio (SNR) $ > 25$. One characteristic of a narrow-band signal that helps to distinguish between fixed i.e. terrestrial transmitters and extraterrestrial transmitters is that the latter are expected to drift in frequency due to the relative motion of the observer and transmitter. However,  since this drift rate is unknown, a large range of trial drift rates must also be applied to the data during the search procedure. 

\cite{enriquez2017} conducted a thorough analysis of the data, employing a Doppler drift rate of $\pm 2$~Hz~s$^{-1}$, and identifying 11 events that were considered to be potential candidates. A more detailed analysis of each event, however, indicated that they were consistent with known  examples of human-generated RFI. They concluded that none of the 692 target stars hosted high duty cycle (continuous) transmitters, operating between 1.1-1.9 GHz with an Equivalent Isotropic Radiated Power (EIRP) in excess of $10^{13}$ Watts. In addition, \cite{enriquez2017} suggested that this implied that less than $\sim 0.1\%$ of stellar systems within 50~pc possess these types of powerful transmitters. 

\citet{Price2020} observed a selection of stars from \citet{isaacson2017} sample with the Green Bank and Parkes radio telescopes, as well as a small number of stars within 5~pc added for improved volume completeness below a declination of \ang{-15}. A total of 1327 primary targets were selected to be observed by the GBT and Parkes radio telescopes. In their analysis, \citet{Price2020} made a substantial improvement on the results of \cite{enriquez2017}, by employing a Doppler drift rate of $\pm 4$~Hz~s$^{-1}$ and lowering the SNR of the original analysis to $> 10$, thus improving the limiting sensitivity (EIRP$_{\rm min}$) of the survey by a factor of 2.5. \cite{Price2020} also increased the number of target stars observed with the GBT at L-band to 882 and added new observations of 1005 stars with the GBT at S-band, and 189 stars with the Parkes Telescope also at S-band. This increased the total number of individual stars observed to 1327. Conducting a similar analysis to \cite{enriquez2017}, they concluded that none of these stars hosted a high duty cycle transmitter with an EIRP in excess of $2.1\times10^{12}$~W at L- and S-band using the GBT or in excess of $9.1\times10^{12}$~W using Parkes. These are by far the most sensitive and thorough SETI observations conducted to date, and represent the first steps towards a final Breakthrough Listen survey goal of observing 1 million of the nearest stars. 

Clearly the number of stars surveyed is an important factor in any SETI survey figure of merit. Traditionally, SETI researchers usually consider observations made by large radio telescopes to be solely targeting individual stars in the centre of the field. This simplifies the analysis but the reality is that a parabolic radio telescope is sensitive to emission across the response of its full primary beam, corresponding to an area of sky that goes well beyond the confines of any particular central target star and includes a large number of background stars (and even a significant number of foreground stars). These are located within the full width at half maximum (FWHM) of the primary beam response of even the largest radio telescopes such as the GBT and Parkes. At the FWHM (full width at half maximum) of a radio telescope beam, the sensitivity of the telescope falls to 50\% of its maximum value at the centre of the beam which is usually synonymous with the pointing centre/target source position. Until now, the impact of the additional foreground and background stars has largely been ignored, partly because the distances to these stars were typically unknown. 

The {\textit{Gaia}} mission \citep{Gaia2016}, designed to perform accurate astrometry of stars in the Milky Way, has enabled a more detailed analysis. Launched in 2013, the main goal of the mission is to create the largest and most precise 3D map of our Galaxy to date, surveying over 1 billion galactic stars in the process \citep{Gaia2016}. With measured parallaxes (and hence inferrable distances) for over 1.3 billion stars in the {\textit{Gaia}} DR2 \citep{Gaia2018summary}, including stars that fall within the field of view of GBT and Parkes observations of Breakthrough Listen targets, it is possible to increase the number of stars formally surveyed in terms of both relatively nearby ($< 50$~pc) and in particular, more distant stellar systems ($> 50$~pc). In short, we can greatly improve limits on the incidence of nearby high duty cycle (continuous) transmitters, as well as those that might be located at much larger distances.  

%The {\textit{Gaia}} mission \citep{Gaia2016} has enabled a more detailed analysis, since many stars that fall within the primary beam of the GBT and Parkes have measured parallaxes and therefore inferred distances. With this information, it is possible to do a much better job, increasing the number of stars formally surveyed in terms of both relatively nearby ($< 50$~pc) and in particular, more distant stellar systems ($> 50$~pc). In short, we can greatly improve limits on the incidence of nearby high duty cycle (continuous) transmitters, as well as those that might be located at much larger distances. 

In this paper, we extend the Breakthrough Listen (BL) analysis of \citet{enriquez2017,Price2020} to include an additional 286998 stars that are located within the FWHM of the primary beams of the GBT L-band, GBT S-band and Parkes 10-cm radio telescope receivers, in addition to 1317 stars from the original target sample. The extended sample of stars was selected such that these have source parallax data included in {\textit{Gaia}} DR2, from which distances can be inferred. Section \ref{2} describes how we define our new and enlarged sample of stars based on an interrogation of the {\textit{Gaia}} archive. We also present some details of the main characteristics of the enlarged sample. Section \ref{3} presents a new analysis of the Continuous Waveform Transmitter Rate Figure of Merit (CWTFM) of \citet{enriquez2017,Price2020}, permitting us to place much stronger limits on the prevalence of powerful transmitters located in our own Galaxy - both within 50~pc and beyond. Grouping stars by their position on the H-R diagram and their effective temperature, we also present an analysis of prevalence of extraterrestrial transmitters by stellar type and spectral class. In Section \ref{4} we summarise the results and draw our conclusions.

\begin{figure*}
\includegraphics[width=\linewidth]{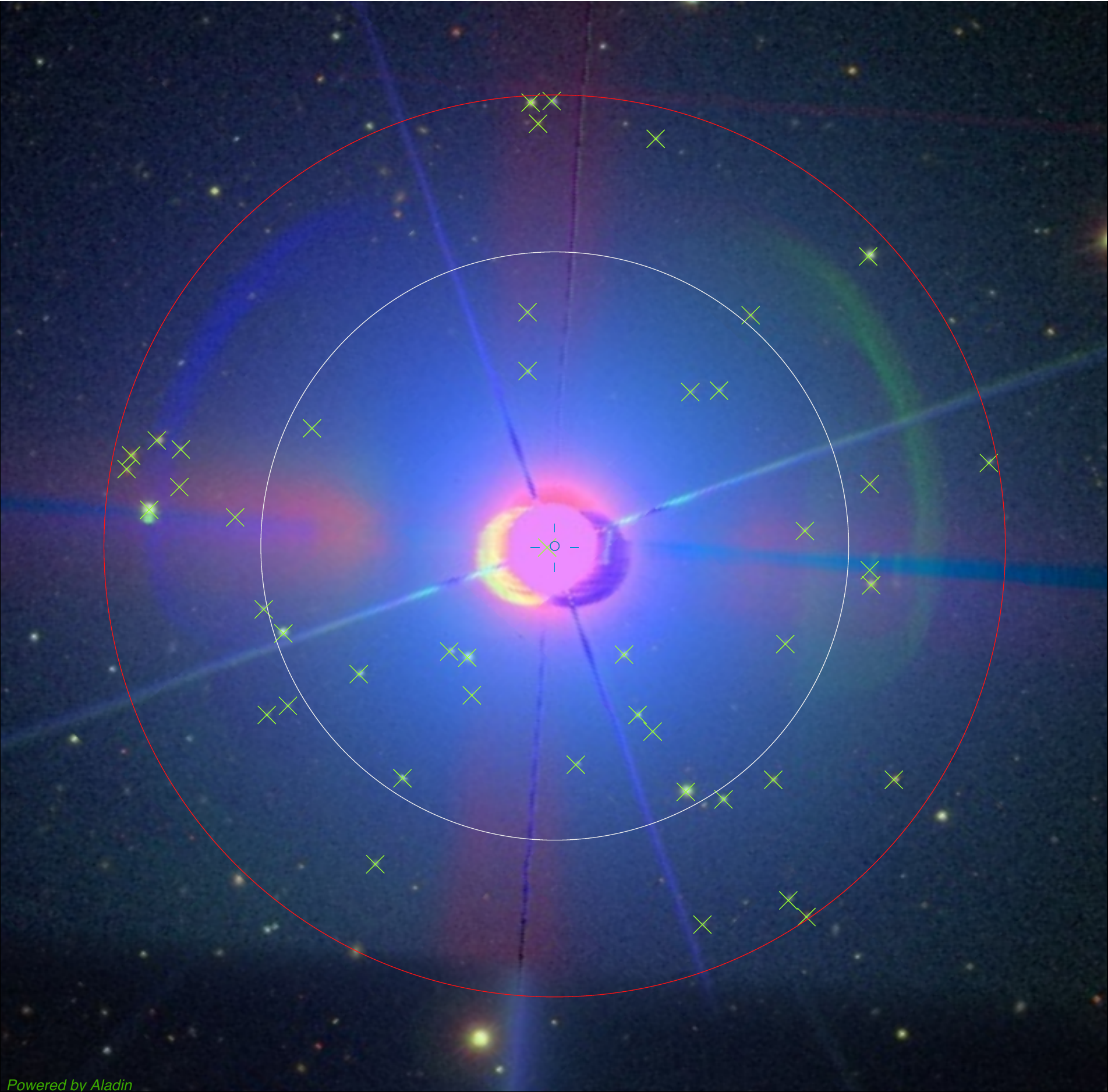}
\caption{An optical colour image of the stellar field centred on HIP109427 from the Pan-STARRS DR1 z and g broadband filters, showing the extent of the FWHM for the GBT L-band and GBT S-band receivers, circled in red and white respectively. 46 sources with geometric distances calculated from {\textit{Gaia}} parallax data are marked with green crosses.} \label{f1}
\end{figure*}

\section{Extending the Breakthrough Listen survey sample} \label{2}

In order to avoid introducing distance-related biases,  we used the external {\textit{Gaia}} DR2 "geometric distance" catalogue presented by \citet{BailerJones2018} to obtain distance estimates (and associated uncertainties) for the sources in our extended sample. The FWHM of the GBT L-band and GBT S-band receivers were estimated to be $\sim 8.4$ and $\sim 5.5$ arc minutes respectively by applying the edge taper calculations presented in the Green Bank Observatory Proposer's Guide\footnote{\url{https://www.gb.nrao.edu/scienceDocs/GBTpg.pdf}}. The FWHM of the Parkes 10-cm receiver was estimated to be 6.4 arc minutes from the Parkes User's Guide\footnote{\url{https://www.parkes.atnf.csiro.au/observing/documentation/users_guide/html/pkug.html}}. Note that the FWHM of the primary beam is dependent on the observing frequency. Since both the GBT and Parkes observe over a finite band, the FWHM presented here and used in our subsequent analysis is calculated for the central frequency of the observations, and represents an average across the band. 

Fig. \ref{f1} presents an optical image of the field around HIP109427, one of the original target stars observed by \citet{Price2020}. In addition to the target at the centre of the field, 45 other stars with distances measured by {\textit{Gaia}} are located within the FWHM of the GBT primary beam at L-band. It should be noted that a few extended objects (likely to be relatively nearby galaxies) also lie within the telescope's field of view; the restriction of our sample to stars neglects the presence of these extended objects, as well as the possibility of "free-floating" transmitters not associated with any particular star.

%\begin{figure*}
%\includegraphics[width=\linewidth]{Distance_magnitude_10kpc_020720.pdf}
%\caption{G-band mean absolute magnitude plotted against source distance for the targets used by \protect\cite{Price2020} (blue crosses) and the corresponding additional sources from our extended sample (marked as WG\&S targets, red crosses), out to a distance of 10~kpc.} \label{f2}
%\end{figure*}

\subsection{{\textit{Gaia}} archive results}
\label{2.1} % used for referring to this section from elsewhere

% The \citet{isaacson2017} Breakthrough Listen target sample was selected by including all stars located within a distance of 5~pc contained within the Gliese and RECONS catalogues, aiming to achieve a broad sampling of stars from the Hipparcos catalogue at distances between 5 and 50~pc. This was achieved by defining domains in B-V colour and V-band magnitude, along the entire length of the main sequence and selecting the 100 nearest stars from each domain. In addition to the main sequence targets, a further 100 stars in the 5-50~pc range were selected from another domain constructed to cover giant and sub-giant stars. The focus on nearby stars in the \citet{isaacson2017} sample allowed relatively low limits to be placed on the power of any transmitters, in accordance with the inverse square law. \citet{Price2020} observed a selection of stars from this sample with the Green Bank and Parkes radio telescopes, as well as a small number of stars within 5~pc added for improved volume completeness below a declination of \ang{-15}. A total of 1327 primary targets were selected to be observed by the GBT and Parkes radio telescopes.

\begin{figure*}
\includegraphics[width=\linewidth]{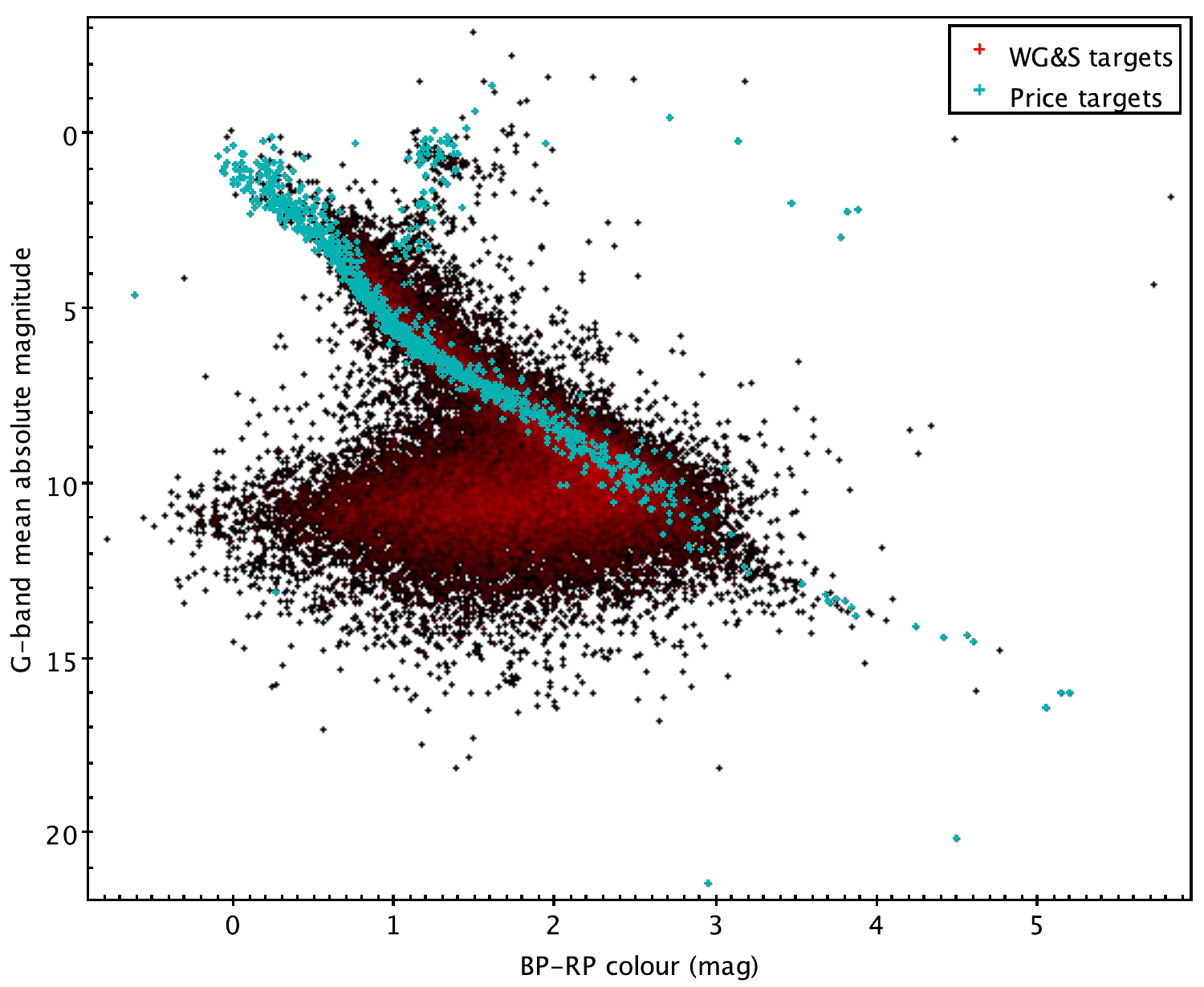}
\caption{Hertzsprung-Russell diagram of the 32683 individual sources in our extended sample at $d<1$~kpc with available G-band absolute magnitude and BP-RP colour values, marked as WG\&S targets. The colour scale represents the square root of the relative density of stars - the points in areas with the fewest stars are black while points in in the most star-dense areas of the diagram are in red. 1228 targets from the original Breakthrough Listen sample of \citet{Price2020} that were cross-matched to a single {\textit{Gaia}} DR2 source are plotted as blue crosses.} \label{f11}
\end{figure*}

To compile our extended sample, the online {\textit{Gaia}} archive was used to probe the {\textit{Gaia}} DR2 \citep{Gaia2018summary} with a list of target field coordinates drawn from the online supplementary tables of \citet{Price2020}. These positions were split across a number of data files in order to prevent client timeout issues associated with the large sample size, and uploaded to the {\textit{Gaia}} archive search engine. We set the search radius to 6.75 arc minutes (well over 1.5 times the FWHM of the GBT at L-band, the receiver with the widest field of view used in the Breakthrough Listen survey to date) in order to have some initial flexibility in defining a complete sample. Stars outside of the FWHMs were later discarded in terms of the analysis presented here. 

After cross-matching the resulting source sample against the geometric distance catalogue presented by \citet{BailerJones2018} and subsequently restricting the sample to include only those stars with an estimated distance of $<10$~kpc, a definitive sample of 288315 unique sources was established (including 1317 stars from the original sample of \citealt{Price2020}) with distances ranging from 1.3012{\raisebox{0.5ex}{\tiny$^{+0.0003}_{-0.0003}$}}-9986.5{\raisebox{0.5ex}{\tiny$^{+3682.5}_{-2610.4}$}}~pc. We note that the extended sample is a factor of $\sim 219\times$ larger than the original sample. In Fig. \ref{f11} an H-R diagram for the extended sample out to a distance of 1~kpc is presented, with the Breakthrough Listen sample plotted on top. The extended sample, presented in Table \ref{T1}, contains a wide variety of stars with different spectral types. 

% We established the $\sfrac{1}{3}$ limit on the fractional distance error through a process of trial and error - attempting to find a compromise between sample size and veracity of the associated stellar distance data. Abandoning limits on the fractional distance error altogether, significantly expands the size of the sample (by a factor of $\sim 6$) but introduces individual sources with poorly defined distances - this is observed in the fidelity of the associated H-R diagrams. Artefacts appear across the H-R diagram, some of which are associated with multiple star systems for which the single-star {\textit{Gaia}} astrometric solution is unreliable \citep{Gaia_HR_2018}. Although a number of these observations represent genuine stars in multiple systems, a majority are in fact artefacts \citep{Arenou2018}. Despite applying no further filters to our data, the form of our extended sample in the H-R plot in Fig. \ref{f11} is very similar to that presented in \citet{Gaia_HR_2018} (their Fig. 1). This suggests that the adoption of the $\sfrac{1}{3}$ limit on the fractional distance error is a good compromise, generating an extended sample that is associated with robust and reliable stellar distance measurements.

%In particular, in order to achieve a compromise between sample completeness and the presence of artefacts in the H-R diagram as shown in Fig. \ref{f11}, in particular between the white dwarfs (shown towards the bottom, left hand side of Fig. \ref{f11}) and the main sequence. 

\begin{table}
\vspace{0.75cm}
\centering
\begin{sideways}
\begin{minipage}{\textheight-1.5cm}
\caption{Truncated Table of the 288315 stars included as part of our extension of the original Breakthrough Listen sample of \citet{Price2020}.}
\newcolumntype{P}[1]{>{\centering\arraybackslash}p{#1}}
\begin{tabularx}{\linewidth}{P{3.3cm}ccP{2cm}ccP{1.3cm}cP{1.2cm}P{2cm}}
%\begin{tabularx}{\textheight}{P{3.3cm}P{1.8cm}P{2cm}P{2cm}P{1.6cm}P{1.25cm}P{1.3cm}P{2cm}P{1.2cm}P{2cm}}
        \toprule
        {\textit{Gaia}} DR2 Source ID & R.A. (deg) & Dec. (deg) & G-band mean absolute magnitude & Target field ID & Observed by & FWHM (arc minutes) & Offset (arc minutes) & Source distance (pc) & EIRP$_{\rm min}$ (W)\\
        \midrule
        703790044252850688 & 127.4502655 & 26.77337094 & 14.42 & GJ1111 & GBT L & 8.4 & 0.359004541 & 3.580 & 1.08E+10\\
        703784061361529984 & 127.4446044 & 26.73860019 & 13.99 & GJ1111 & GBT L & 8.4 & 2.340012752 & 139.3 & 2.02E+13\\
        703790005596367104 & 127.4737523 & 26.78528586 & 11.30 & GJ1111 & GBT L & 8.4 & 1.221205851 & 268.7 & 6.43E+13\\
        703790009893110400 & 127.4745381 & 26.78483407 & 8.847 & GJ1111 & GBT L & 8.4 & 1.251640544 & 273.5 & 6.68E+13\\
        $\cdots$ & & & & & & & & &\\
        1856712201709083136 & 307.5436950 & 26.84226469 & 7.860 & HIP101150 & GBT S & 5.5 & 0.056771902 & 20.40 & 3.50E+11\\
        1856712236068832256 & 307.5192821 & 26.84375964 & 8.326 & HIP101150 & GBT S & 5.5 & 1.511165923 & 300.4 & 9.35E+13\\
        1856712197402307712 & 307.5429722 & 26.83802543 & 9.962 & HIP101150 & GBT S & 5.5 & 0.301823444 & 342.6 & 9.94E+13\\
        1856713438650729600 & 307.5899697 & 26.84265686 & 10.50 & HIP101150 & GBT S & 5.5 & 2.731136293 & 483.6 & 3.89E+13\\
        $\cdots$ & & & & & & & & &\\
        5698188160215537920 & 119.5199403 & -25.62768698 & 6.730 & HIP38939 & Parkes & 6.4 & 0.121689109 & 18.47 & 1.24E+12\\
        5698199876886342144 & 119.4882756 & -25.61820767 & 15.49 & HIP38939 & Parkes & 6.4 & 1.865795860 & 111.8 & 5.76E+13\\
        5698194104450235392 & 119.5583322 & -25.60731470 & 5.851 & HIP38939 & Parkes & 6.4 & 2.671703306 & 179.3 & 1.90E+14\\
        5698187743587483008 & 119.5232805 & -25.67715221 & 11.59 & HIP38939 & Parkes & 6.4 & 3.046661268 & 216.5 & 3.20E+14\\
        \bottomrule
        \label{T1}
\end{tabularx}
\end{minipage}
\end{sideways}
\end{table}

\subsection{Characteristics of the extended sample}
\label{2.2}

Clearly our extended analysis adds many stars of different spectral type to the original Breakthrough Listen sample (see Fig. \ref{f11}). The original sample is narrowly clustered around the main sequence and the red giant branch. The extended sample greatly expands the number of main sequence and red giant stars but also adds significant numbers of other spectral types. In Fig. \ref{f11}, hot luminous stars are shown in the upper left of the figure and cooler faint stars located further towards the bottom right. The giant branch is well populated by the extended sample, and there are a number of white dwarfs also present. 

%As we shall see later, grouping stars by their location on the H-R plot as well as knowing???? their spectral cpermits us to also draw conclusions about the prevalence of transmitting civilisations based on their stellar type. 

% \begin{figure*}
% \includegraphics[width=\linewidth]{Distance_magnitude_1kpc_020720.pdf}
% \caption{G-band mean absolute magnitude plotted against source distance for the targets used by \protect\cite{Price2020} (blue crosses) and the corresponding additional sources from our extended sample (marked as WG\&S targets, red crosses), out to a distance of 1~kpc.} \label{f6}
% \end{figure*}

\begin{figure*}
\includegraphics[width=\linewidth]{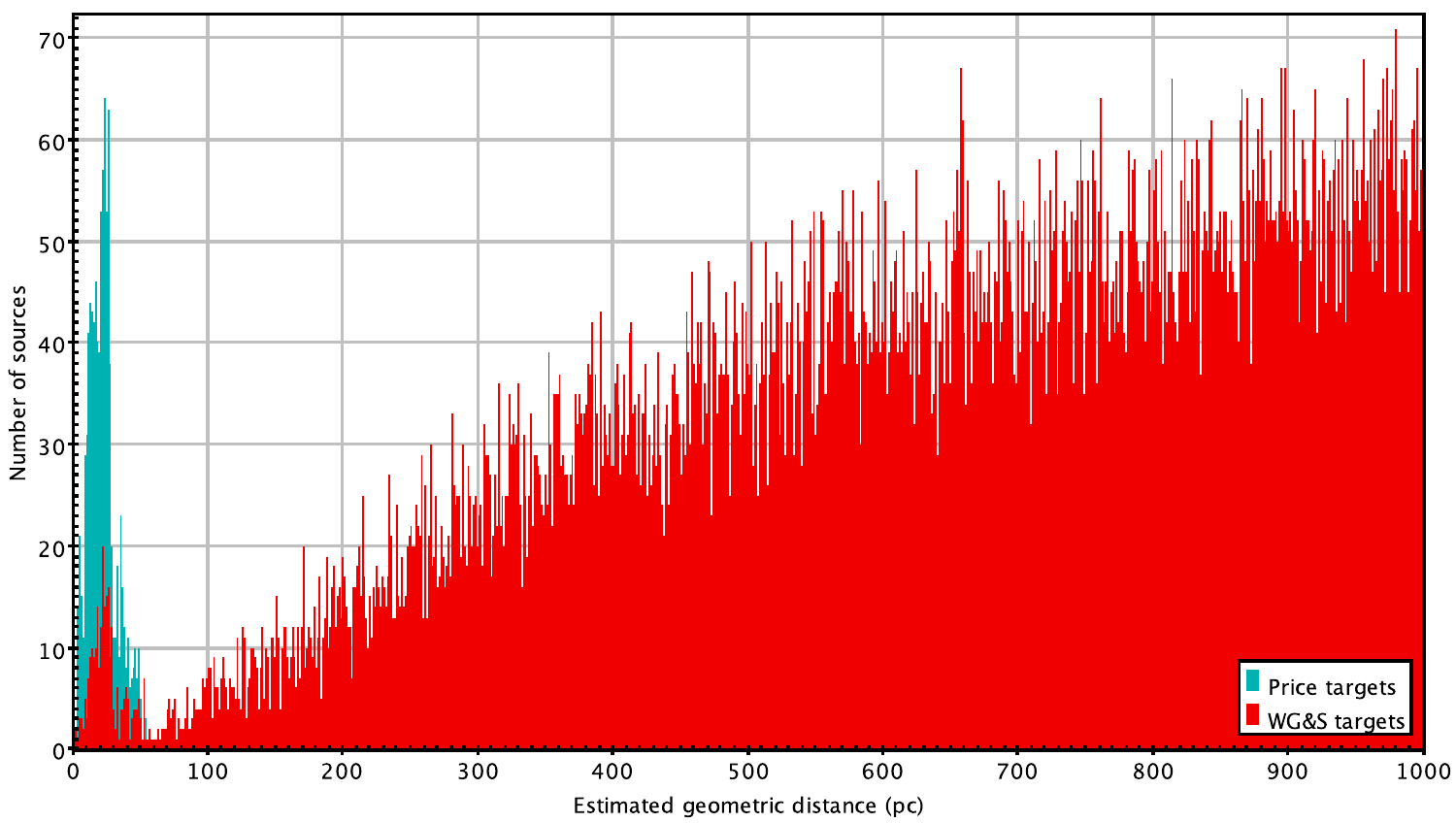}
\caption{Number of sources in each 1~pc distance bin plotted against source distance for the targets used by \protect\cite{Price2020} and the corresponding additional sources from our extended sample (marked as WG\&S targets), out to a distance of 1~kpc.}\label{f4}
\end{figure*}

We were initially surprised that not all of the Breakthrough Listen stellar sample have cross-matched sources in the {\textit{Gaia}} DR2 catalogue, even after a careful manual analysis. It turns out, that while the {\textit{Gaia}} survey is essentially complete for apparent magnitudes between ${\rm G} = 12$ and ${\rm G} = 16$, there is a reduction in completeness for brighter stars with ${\rm G} \lesssim 7$, and that most stars with ${\rm G} \lesssim 3$ are missing from the {\textit{Gaia}} DR2 catalogue \citep{Gaia2018parallax}.  Similarly, \citet{Gaia2018parallax} further acknowledge that around 20\% of stars with proper motions $\gtrsim 0.6$~\si{arcsec.yr^{-1}} also remain missing from the catalogue. Since the \citet{Price2020} sample is populated with bright, nearby stars, often with large proper motions, 75 targets from their sample have no counterpart in the {\textit{Gaia}} DR2 catalogue. In addition, \citet{isaacson2017} chose to regard  binary systems with separation $< 2$~arcsec as single targets in their sample, in addition to other double and multiple stars that were considered to constitute a single target pointing in the samples of both \citet{enriquez2017} and \citet{Price2020}. However, {\textit{Gaia}} spatially resolves many of these systems and this leads to issues in directly cross-matching a number of targets from \citet{Price2020} to singular and unique  counterparts in the {\textit{Gaia}} DR2 catalogue. Only those targets from \citet{Price2020} that could be directly matched to a single {\textit{Gaia}} DR2 source are plotted as "Price targets" in the figures presented in this section. 

Fig. \ref{f4} presents a histogram showing the number of sources in each 1~pc bin, out to a distance of 1~kpc. A relatively high density of stars at distances of $<50$~pc is seen in the extended sample, since as noted earlier, many of the original targets are actually nearby binary or multiple stellar systems. Excluding the large number of stars at distances $<50$~pc, an apparent under-density of sources at distance scales $\sim 50-100$~pc is also seen in the extended sample. This under-density is also present in the {\textit{Gaia}} sample as a whole at distances $\lesssim 100$~pc and reflects the aforementioned limitations in the {\textit{Gaia}} sample completeness for nearby sources that are typically bright, often with large proper motions \citep{Gaia2018parallax}.

There is one source, HIP1692, included in the \citet{isaacson2017} 5-50~pc sample which also featured in the sample used by \citet{Price2020}. They assume a distance of 23.0~pc, a value calculated from its parallax as presented in the original 1997 Hipparcos main catalogue \citep{Hipparcos1997}. However, both the 2007 Hipparcos "new reduction" catalogue \citep{Hipparcos2007} and the data from {\textit{Gaia}} DR2, revealed it to be much further away, with 
parallaxes quoted for the Hipparcos and {\textit{Gaia}} data sets resulting in distances of 310.56~pc and 769.37~pc respectively, while the geometric distance calculation of \citet{BailerJones2018} reports a distance of $\sim 752.61${\raisebox{0.5ex}{\tiny$^{+28.91}_{-26.88}$}}~pc. We kept this target in our extended sample with the estimated geometric distance from \citet{BailerJones2018} assumed.

%it can be clearly seen in Fig. \ref{f6} as an outlier of the \citet{Price2020} sample.

%\begin{table}
%	\centering
%	\caption{This is an example table. Captions appear above each table.
%	Remember to define the quantities, symbols and units used.}
%	\label{tab:example_table}
%	\begin{tabular}{lccr} % four columns, alignment for each
%		\hline
%		A & B & C & D\\
%		\hline
%		1 & 2 & 3 & 4\\
%		2 & 4 & 6 & 8\\
%		3 & 5 & 7 & 9\\
%		\hline
%	\end{tabular}
%\end{table}

\section{Results and Discussion} \label{3}

Drawing from our extended sample, we add 286998 stars to the sample of 1317 targets taken from the original analysis of \citet{enriquez2017} and \citet{Price2020}. This increases the number of stars observed within a distance of 50~pc of the Earth by 196{\raisebox{0.5ex}{\tiny$^{+42}_{-7}$}}. Uncertainty in the number of stars reflects uncertainty in the geometric distance estimates of \citet{BailerJones2018}. Our extended sample of 286802{\raisebox{0.5ex}{\tiny$^{+7}_{-42}$}} stars with distances in excess of 50~pc opens up a new range of analysis space for the Breakthrough Listen survey. Both samples permit us to extend the original limits placed by \citet{enriquez2017} and \citet{Price2020} on the incidence of powerful transmitters associated with both nearby and more distant stellar systems.

The Equivalent Isotropic Radiated Power (EIRP, in W) generated by another civilisation depends on the power of the transmitter, $P_{\rm tx}$, and its antenna gain, $G_{\rm ant}$, such that:

\begin{equation}
{\rm EIRP}=G_{\rm ant} P_{\rm tx}.
\end{equation}

The minimum EIRP that can be detected is then: 
\begin{equation}
{\rm EIRP}_{\rm min}  = 4\pi d^{2}{F_{\rm min}}, 
\end{equation}

where $d$ is the distance to the transmitter, and $F_{\rm min}$ is the minimum flux (in units of~W/m$^2$) that can be detected by the observing system. The minimum detectable flux $F_{\rm min} = S_{\rm min} \delta\nu_{t}$, where $S_{\rm min}$ is minimum detectable flux density and $\delta\nu_{t}$ is the bandwidth of a transmitted signal; a bandwidth value of unity has been assumed for this work \citep{Price2020}. At a distance of 50~pc, the EIRP$_{\rm min}$ for the GBT L-band and S-band receivers is $2.1\times10^{12}$~W, and for the Parkes 10-cm receiver it is $9.1\times10^{12}$~W \citep{Price2020}.

We calculated the EIRP$_{\rm min}$ for each source in our extended sample by scaling the calculations of \citet{Price2020}, taking into account both source distance and the relative position of the source in the beam. We adopted a Gaussian function for the changing response of the telescope across the field. The EIRP$_{\rm min}$ associated with each star therefore takes into account not only its distance but also its offset from the telescope pointing position (centred on the original target star). Table \ref{T1} presents a representative list of the extended source sample and associated {\textit{Gaia}} measurements and EIRP$_{\rm min}$ values\footnote{The full table is available on the online version of this paper.}. 

\cite{enriquez2017} suggested a new figure of merit for SETI surveys, and for the Breakthrough Listen Initiative in particular. This attempts to take into account the limitations of earlier figures of merit, and includes all the parameters relevant to the final Breakthrough Listen 1 million star survey. More specifically, \cite{enriquez2017} and \citet{Price2020} define a Continuous Waveform Transmitter Rate Figure of Merit (CWTFM) such that:

\begin{equation}
    {\rm CWTFM} = \zeta_{AO}~\frac{{\rm EIRP_{min}}}{N_{\rm stars}\nu_{\rm rel}},
\end{equation}

where $N_{\rm stars}$ is the total number of stars observed and the fractional bandwidth $\nu_{\rm rel}$ is the total survey bandwidth, $\Delta\nu_{\rm tot}$, divided by the central frequency, $\nu_{\rm mid}$. The normalisation factor $\zeta_{\rm AO}$ is based on the performance of the Arecibo Planetary Radar system, such that $\rm CWTFM = 1$ when ${\rm EIRP} = L_{\rm AO} = 10^{13}$~W, $\nu_{\rm rel} = 1/2$ and $N_{\rm stars} = 1000$. CWTFM scores with smaller values (e.g. $<1$), represent surveys that are more complete and/or have better sensitivity. 

%In addition to the aforementioned limiting of artefacts in the H-R diagram in Fig. \ref{f11}, as the derived EIRP$_{\rm min}$ depends on the square of the distance to the source, adopting a limiting fractional distance error of $\sfrac{1}{3}$ for sources in our extended sample also serves to limit the uncertainty on derived EIRP$_{\rm min}$ values. 

The initial survey of \cite{enriquez2017} has a CWTFM $\sim 0.85$, while the more extensive surveys of \cite{Price2020} yield CWTFMs of $\sim 0.11$ for the GBT L- and S-band observations and $\sim 8.21$ for the Parkes observations. Note that lower CWTFM values imply a better figure of merit. For comparison, the well known Project Phoenix survey \citep{Backus2002} has a CWTFM of $\sim 49$ \citep{enriquez2017}.  

\subsection{Results for the extended sample}
\label{3.1}

While the EIRP$_{\rm min}$ is a simple function of stellar distance for \citet{enriquez2017} and \citet{Price2020}, with our approach it depends not only on distance but also the offset of any given star from the centre of the primary beam. With a large sample of stars spanning a wide range of distances and different offsets, it makes more sense to group the stars in shells of increasing EIRP$_{\rm min}$. Grouping the stars in such shells, allows us to obtain CWTFMs of $\sim$ 0.0551{\raisebox{0.5ex}{\tiny$^{+0.0001}_{-0.0015}$}} for the GBT observations and $\sim$ 1.38{\raisebox{0.5ex}{\tiny$^{+0}_{-0.07}$}} for the Parkes observations (both for the shells with EIRP$_{\rm min} = 10^{12}$~W). In Table \ref{T2}, we present the CWTFM figures for our analysis for a number of different EIRP$_{\rm min}$ shells. Uncertainties in the calculated CWTFM values are also presented, as calculated from the upper and lower bounds on the confidence interval of the estimate distance values presented in the \citet{BailerJones2018} geometric distance catalogue.

\begin{table}
\centering
\caption{CWTFM figures for the analyses of \citet{enriquez2017} and \citet{Price2020}, together with our figures for extended sample calculated over a range of EIRP$_{\rm min}$ shells.}
\begin{tabular}{cccc}
    \toprule
    & EIRP$_{\rm min}$ & $N_{\rm *}$ & CWTFM\\
    \midrule
    Enriquez GBT & $5.2 \times 10^{12}$~W & 692 & 0.85\\
    Price GBT & $2.1 \times 10^{12}$~W & 1213 & 0.11\\
    Price Parkes & $9.1 \times 10^{12}$~W & 189 & 8.21\\
    \midrule
    \multirow{8}{*}{WG\&S GBT} & $1 \times 10^{11}$~W & 145{\raisebox{0.5ex}{\tiny$^{+27}_{-1}$}} & 0.0448{\raisebox{0.5ex}{\tiny$^{+0.0003}_{-0.0070}$}}\\
    & $1 \times 10^{12}$~W & 1180{\raisebox{0.5ex}{\tiny$^{+32}_{-3}$}} & 0.0551{\raisebox{0.5ex}{\tiny$^{+0.0001}_{-0.0015}$}}\\
    & $1 \times 10^{13}$~W & 1491{\raisebox{0.5ex}{\tiny$^{+86}_{-14}$}} & 0.436{\raisebox{0.5ex}{\tiny$^{+0.004}_{-0.024}$}}\\
    & $1 \times 10^{14}$~W & 3736{\raisebox{0.5ex}{\tiny$^{+2427}_{-650}$}} & 1.74{\raisebox{0.5ex}{\tiny$^{+0.37}_{-0.69}$}}\\
    & $1 \times 10^{15}$~W & 26384{\raisebox{0.5ex}{\tiny$^{+33570}_{-11433}$}} & 2.46{\raisebox{0.5ex}{\tiny$^{+1.88}_{-1.38}$}}\\
    & $1 \times 10^{16}$~W & 149518{\raisebox{0.5ex}{\tiny$^{+79977}_{-76733}$}} & 4.35{\raisebox{0.5ex}{\tiny$^{+4.58}_{-1.51}$}}\\
    & $1 \times 10^{17}$~W & 249734{\raisebox{0.5ex}{\tiny$^{+256}_{-7807}$}} & 26.03{\raisebox{0.5ex}{\tiny$^{+0.84}_{-0.03}$}}\\
    & $1 \times 10^{18}$~W & 249990{\raisebox{0.5ex}{\tiny$^{+0}_{-0}$}} & 260.01{\raisebox{0.5ex}{\tiny$^{+0}_{-0}$}}\\
    \midrule
    \multirow{8}{*}{WG\&S Parkes} & $1 \times 10^{11}$~W & 19{\raisebox{0.5ex}{\tiny$^{+7}_{-1}$}} & 0.90{\raisebox{0.5ex}{\tiny$^{+0.05}_{-0.24}$}}\\
    & $1 \times 10^{12}$~W & 124{\raisebox{0.5ex}{\tiny$^{+7}_{-0}$}} & 1.38{\raisebox{0.5ex}{\tiny$^{+0}_{-0.07}$}}\\
    & $1 \times 10^{13}$~W & 207{\raisebox{0.5ex}{\tiny$^{+9}_{-0}$}} & 8.24{\raisebox{0.5ex}{\tiny$^{+0}_{-0.34}$}}\\
    & $1 \times 10^{14}$~W & 272{\raisebox{0.5ex}{\tiny$^{+62}_{-25}$}} & 62.72{\raisebox{0.5ex}{\tiny$^{+6.35}_{-11.64}$}}\\
    & $1 \times 10^{15}$~W & 1129{\raisebox{0.5ex}{\tiny$^{+1131}_{-394}$}} & 151.10{\raisebox{0.5ex}{\tiny$^{+81.00}_{-75.62}$}}\\
    & $1 \times 10^{16}$~W & 7564{\raisebox{0.5ex}{\tiny$^{+11701}_{-4074}$}} & 225.53{\raisebox{0.5ex}{\tiny$^{+263.27}_{-136.98}$}}\\
    & $1 \times 10^{17}$~W & 41612{\raisebox{0.5ex}{\tiny$^{+10792}_{-24101}$}} & 409.95{\raisebox{0.5ex}{\tiny$^{+564.23}_{-84.42}$}}\\
    & $1 \times 10^{18}$~W & 53614{\raisebox{0.5ex}{\tiny$^{+0}_{-67}$}} & 3181.79{\raisebox{0.5ex}{\tiny$^{+3.98}_{-0}$}}\\
    \bottomrule
    \label{T2}
\end{tabular}
\end{table}

\begin{figure*}
\centering
\includegraphics[width=\linewidth]{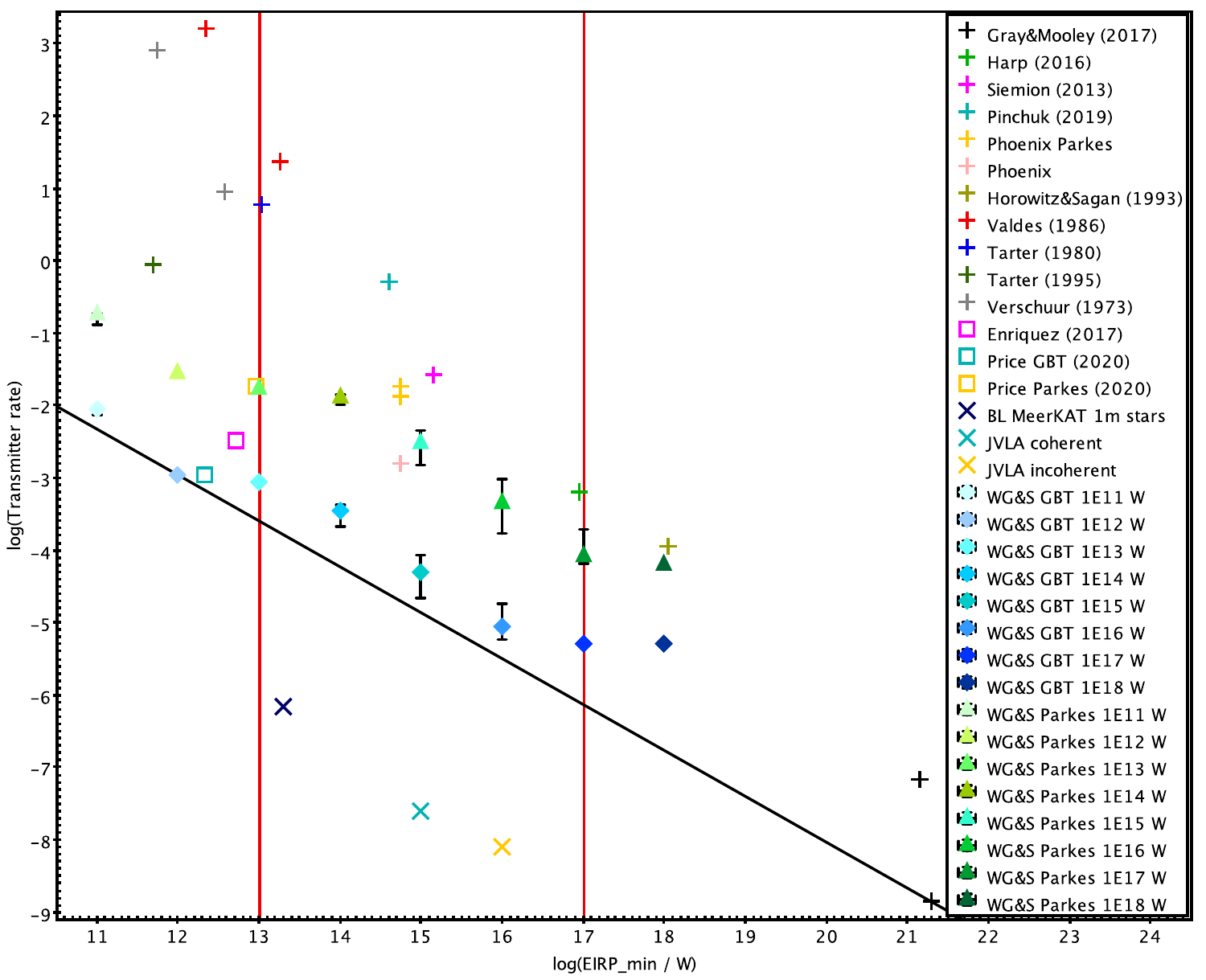}
\caption{Transmitter rate plotted on logarithmic axes against the EIRP$_{\rm min}$ for a range of historical projects, including the Breakthrough Listen Initiative (after \protect\citealt{Price2020}), as well as our extended sample organised in shells of increasing EIRP$_{\rm min}$ (marked by points labelled "WG\&S", along with the telescope they were observed by, and the maximum EIRP$_{\rm min}$ of sources in the shell). The uncertainty in transmitter rate for each shell is also shown. For comparison, we also include projected values for the Breakthrough Listen 1 million star survey using MeerKAT, as well as projected values for two VLA VLASS commensal SETI observing modes from \citet{Hickish2019}. The black line is a fit of the values for this work and that of \citet{Gray2017}. The vertical lines at
${\rm log}({\rm EIRP}_{\rm min}/{\rm W}) = 13$ and ${\rm log}({\rm EIRP}_{\rm min}/{\rm W}) = 17$ represent the EIRP of the Arecibo planetary radar and the total solar power incident on Earth (commonly referred to as the energy usage of a Kardashev Type I civilisation) respectively.} \label{f12}
\end{figure*}

Following \cite{enriquez2017}, we compare the CWTFM of past SETI surveys to the Breakthrough Listen survey, and its extension as described in this paper. In particular, in Fig. \ref{f12} we plot each survey’s EIRP$_{\rm min}$ versus $(N_{\rm stars}\nu_{\rm rel})^{-1}$, or as \cite{enriquez2017} refer to it, the Transmitter Rate. The uncertainty in transmitter rate values calculated for shells in our analysis is also shown. As can be seen in Fig. \ref{f12}, our approach to extending the sample by including other stars in the field of view of each of the radio telescopes, significantly improves the limits to be placed on the incidence of powerful transmitters (see again Fig.~\ref{f12}) associated with both nearby and distant stellar systems.

Both \cite{enriquez2017} and \cite{Price2020} find no evidence for continuous (100\% duty cycle) transmitters associated with the nearby ($d < 50$~pc) star systems observed. This includes directional transmitters (e.g. radio beacons) directed at the Earth with a power output equal to or greater than the brightest human-made transmitters (e.g. a canonical Arecibo planetary radar-like system with a gain of 70~dB and a transmitter power of $\sim 1$~MW). To detect a non-directional isotropically radiating antenna, the transmitter power must be $\sim 10^{13}$~W  (around the current energy consumption of our own civilisation). From their analysis of 692 stars, \cite{enriquez2017} conclude that fewer than $\sim 0.1$\% of the stellar systems within 50~pc possess these types of transmitters. Our extension of the Breakthrough Listen survey expands the number of stars within 50~pc surveyed for transmitters with an EIRP $\geqslant 10^{13}$~W to a total of 1513{\raisebox{0.5ex}{\tiny$^{+9}_{-7}$}}, slightly improving upon this figure to $\sim 0.0660${\raisebox{0.5ex}{\tiny$^{+0.0004}_{-0.0003}$}}\%.

However, moving beyond the nearby stellar sample, the total number of stars in our extended sample increases significantly (see also Fig. \ref{f4}). This partially compensates for the decrease in survey sensitivity at these larger distances, permitting us to place very interesting limits on civilisations with powerful transmitters, surpassing previous studies quite significantly. For example, we conclude that at 100 and 200~pc, the incidence of star systems with such transmitters is $\lesssim$ 0.061{\raisebox{0.5ex}{\tiny$^{+0.001}_{-0.003}$}}\% (for an EIRP $\gtrsim 6.5 \times 10^{13}$~W), and $\lesssim$ 0.039{\raisebox{0.5ex}{\tiny$^{+0.004}_{-0.008}$}}\% (for an EIRP $\gtrsim 2.5 \times 10^{14}$~W) respectively. 
We note however, that at distances beyond a few hundred pc, interstellar scintillation (ISS) effects may begin to play a role in the Breakthrough Listen analysis. In particular, the expected broadening of very narrow-band signals will make them more difficult to detect at the lowest observing frequencies as they become washed-out across the observing band e.g. see \citet{Cordes1997,Siemion2013}. In addition, scintillation can increase but also decrease the signal's amplitude, making it intermittent on relatively short timescales. Since these effects typically scale with $\nu^{-2}$, the higher frequency S-band data should not be so badly affected. In any case, at frequencies of $\sim 1$~GHz our results beyond 200~pc, i.e. for our shells with ${\rm EIRP}_{\rm min} \geqslant 10^{14}$~W should be treated with some caution. These are the first shells (in order of increasing ${\rm EIRP}_{\rm min}$) for which stellar objects at distances $\gtrsim 200$~pc begin to form a substantial fraction of the total sample. We note that any broadband extraterrestrial signals should not be badly affected by ISS effects. 

\begin{figure*}
    \centering
    \includegraphics[width=\linewidth]{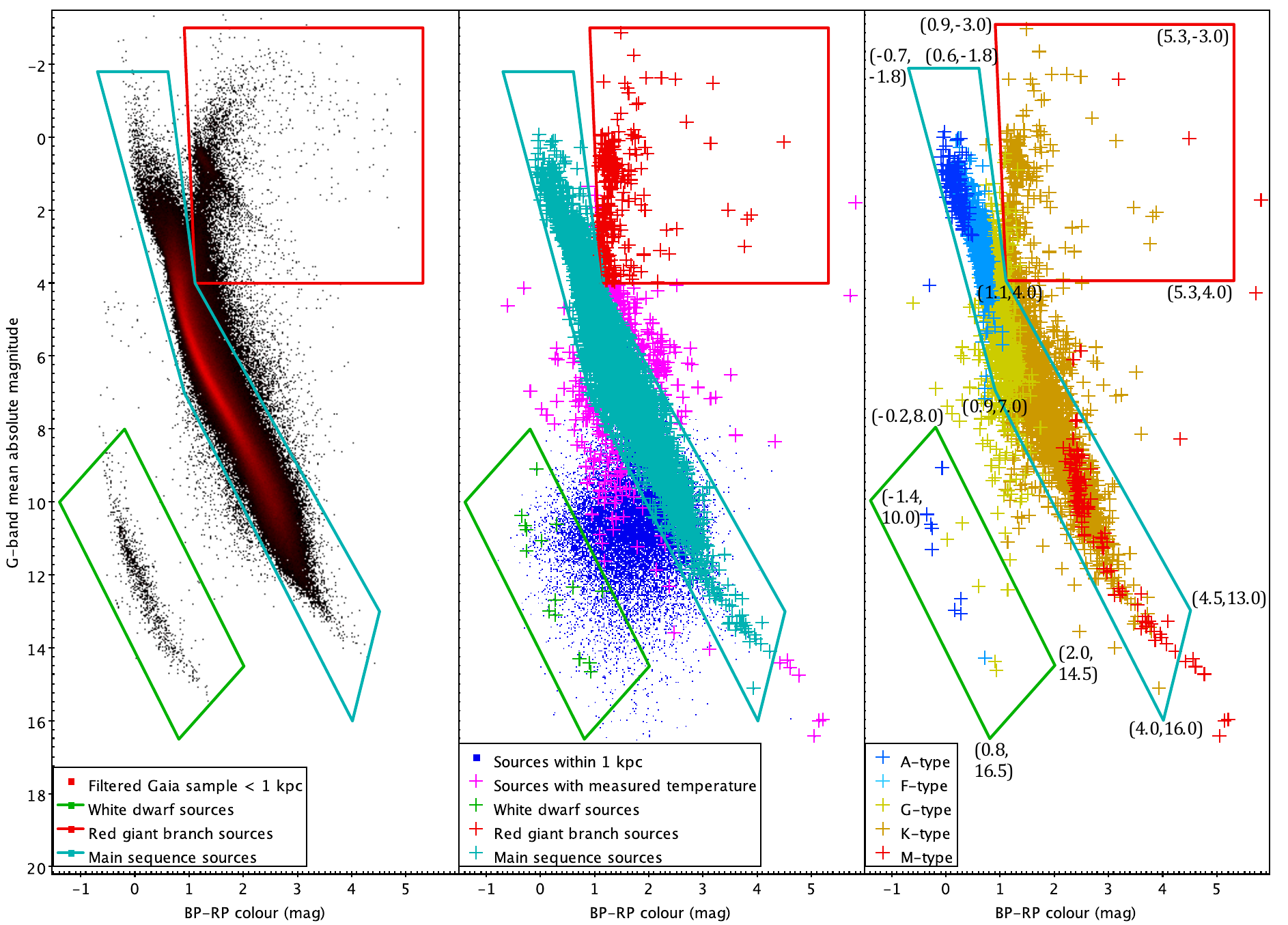}
    \caption{How sources in our extended sample were grouped in order to make estimates of the prevalence of extraterrestrial transmitters associated with sources of different types and spectral classes. Left panel: a filtered sample of random {\textit{Gaia}} sources within 1~kpc used to define domains for white dwarfs, main sequence and red giant branch stars. These are highlighted in green, cyan and red respectively. Middle panel: blue dots mark our extended sample at distances $< 1$~kpc, while crosses represent sources with measured {\textit{Gaia}} DR2 effective temperature values from which spectral class can be calculated. Pink crosses represent sources with measured effective temperature that were excluded from the analysis. Right panel: sources with measured effective temperature are grouped by spectral class.} \label{f13}
\end{figure*}

\subsection{Limits on transmitting civilisations on the basis of stellar type}
\label{3.2}

The inclusion of non-main-sequence stars in our analysis, permits us to set new limits on the prevalence of extraterrestrial transmitters associated with different stellar type.  We collected a random sample of all {\textit{Gaia}} DR2 sources within 1~kpc and applied the filtering scheme presented in Section 2.1 of \citet{Gaia_HR_2018}. This filtering scheme permits the study of fine structure in the H-R diagram at the expense of completeness, allowing us to more accurately define regions containing various types of stars than would be the case if we had used our extended sample. This was due to the large number of sources in our extended sample falling between the white dwarf and main sequence regions (see Fig. \ref{f11}), suggesting the presence of artefacts \citep{Arenou2018} and blurring the boundaries between the regions. Using the random sample as a guide, we constructed domains in BP-RP colour and absolute magnitude defined to broadly contain main sequence stars, red giant branch stars and white dwarfs (see left panel of Fig. \ref{f13}), allowing us to bin sources in our extended sample by type (Fig. \ref{f13}, middle panel). We subsequently used {\textit{Gaia}} effective temperature data to bin the sources by spectral class (Fig. \ref{f13}, right panel).

In Table \ref{T3}, we place limits on the prevalence of transmitting civilisations by stellar type and spectral class, at distances of 200~pc and 1~kpc. These limits are particularly weak for white dwarfs and red giant branch stars, due to the small number of these objects featuring in our extended sample at distances $< 1$~kpc, as well as the lack of {\textit{Gaia}} effective temperature data for a large number of sources falling within the white dwarf domain on the H-R plot (see middle panel, Fig. \ref{f13}). While disregarding spectral classes and grouping sources purely by domain produces much better prevalence limits, particularly for white dwarfs and main sequence stars, it should be noted that our extended sample contains a large number of sources which would be excluded after applying the filtering scheme from \citet{Gaia_HR_2018}, especially in the region of the H-R plot between the white dwarfs and the main sequence. Many of these filtered sources fall within the defined white dwarf and main sequence domains and negligible numbers in the red giant branch (see again middle panel, Fig. \ref{f13}). This may suggest that a significant fraction of these sources are artefacts \citep{Arenou2018} and thus prevalence results for sources purely based on their location in the H-R diagram should be taken with some caution, particularly for white dwarf and main sequence domain sources. Nevertheless, this approach of placing constraints on the basis of stellar class is instructive, and is likely to be a fairly standard product of future SETI surveys that take into account both foreground and background objects in the telescope's field of view.

We also note that white dwarfs represent an interesting target for SETI studies \citep{Gertz2020}. While it seems counter-intuitive that a stellar system at the end of its life cycle might provide conditions that are hospitable to life, in the cooling phase these stars can maintain relatively benign environments that are expected to be stable on time scales of many billions of years. Most recently, observations also show evidence that debris disks and planets can still be found in these evolved systems e.g. \citep{Kozakis2020}. Such features might make white dwarfs potentially interesting targets for SETI programmes. We very much prefer this open-ended approach to SETI surveys, as opposed to highly focused approaches centred on familiar solar-type stars.

%The inclusion of non-main sequence stars in our analysis allows us to also make estimates of the prevalence of transmitting civilisations as a function of spectral type. To compile a list of white dwarfs in our extended sample (without the $<\sfrac{1}{3}$ fractional error requirement), we cross-matched it against the {\textit{Gaia}} DR2 catalogue of white dwarfs presented by \citet{GentileFusillo2019}. Lists of main sequence and red giant branch stars were compiled by their approximate position on the H-R diagram in Fig. \ref{f11}. This allowed us to place weak limits on the prevalence of transmitting civilisations, at limiting distances of 50~pc, 200~pc and 500~pc; these limits, presented in Table \ref{T3}, are particularly weak for white dwarfs and red giant branch stars due to the low number of these objects featuring in our extended sample at distances $< 500$~pc. Selecting a greater number of white dwarfs and RGB stars as targets for future SETI surveys would greatly improve on these estimates.

\begin{table*}
\centering
\caption{Prevalence figures for extraterrestrial transmitters in our extended sample as a function of distance and stellar type. The top row for each domain denotes all sources within that domain, including sources both with and without available effective temperature information (see ``Sources within 1~kpc", middle panel, Fig. \ref{f13}).}
\makebox[\textwidth]{%
    \begin{tabular}{ccccccccc}
        \toprule
        \multirow{2}{*}{Domain} & \multirow{2}{*}{Spectral class} & \multicolumn{3}{c}{200~pc} & & \multicolumn{3}{c}{1~kpc}\\
        & & $N_{*}$ & Prevalence & EIRP$_{\rm min}$ & & $N_{*}$ & Prevalence & EIRP$_{\rm min}$\\
        \midrule
        \multirow{4}{*}{White dwarfs} & - & 208{\raisebox{0.5ex}{\tiny$^{+252}_{-96}$}} & $<$0.5{\raisebox{0.5ex}{\tiny$^{+0.4}_{-0.3}$}}\% & $2.4 \times 10^{14}$~W & & 2005{\raisebox{0.5ex}{\tiny$^{+675}_{-1448}$}} & $<$0.05{\raisebox{0.5ex}{\tiny$^{+0.13}_{-0.01}$}}\% & $7.1 \times 10^{15}$~W\\
        & A & 7{\raisebox{0.5ex}{\tiny$^{+0}_{-0}$}} & $<$14.3{\raisebox{0.5ex}{\tiny$^{+0}_{-0}$}}\% & $1.7 \times 10^{14}$~W & & 8{\raisebox{0.5ex}{\tiny$^{+0}_{-0}$}} & $<$12.5{\raisebox{0.5ex}{\tiny$^{+0}_{-0}$}}\% & $1.9 \times 10^{14}$~W\\
        & F & 1{\raisebox{0.5ex}{\tiny$^{+0}_{-0}$}} & $<$100.0{\raisebox{0.5ex}{\tiny$^{+0}_{-0}$}}\% & $6.6 \times 10^{11}$~W & & 1{\raisebox{0.5ex}{\tiny$^{+0}_{-0}$}} & $<$100.0{\raisebox{0.5ex}{\tiny$^{+0}_{-0}$}}\% & $6.6 \times 10^{11}$~W\\
        & G & 6{\raisebox{0.5ex}{\tiny$^{+0}_{-0}$}} & $<$16.7{\raisebox{0.5ex}{\tiny$^{+0}_{-0}$}}\% & $4.7 \times 10^{13}$~W & & 6{\raisebox{0.5ex}{\tiny$^{+0}_{-0}$}} & $<$16.7{\raisebox{0.5ex}{\tiny$^{+0}_{-0}$}}\% & $4.7 \times 10^{13}$~W\\
        \midrule
        \multirow{6}{*}{Main sequence} & - & 1931{\raisebox{0.5ex}{\tiny$^{+51}_{-44}$}} & $<$0.052{\raisebox{0.5ex}{\tiny$^{+0.001}_{-0.001}$}}\% & $2.5 \times 10^{14}$~W & & 19011{\raisebox{0.5ex}{\tiny$^{+4265}_{-5131}$}} & $<$0.005{\raisebox{0.5ex}{\tiny$^{+0.002}_{-0.001}$}}\% & $7.2 \times 10^{15}$~W\\
        & A & 87{\raisebox{0.5ex}{\tiny$^{+0}_{-0}$}} & $<$1.1{\raisebox{0.5ex}{\tiny$^{+0}_{-0}$}}\% & $1.8 \times 10^{14}$~W & & 126{\raisebox{0.5ex}{\tiny$^{+4}_{-1}$}} & $<$0.79{\raisebox{0.5ex}{\tiny$^{+0.01}_{-0.02}$}}\% & $3.2 \times 10^{15}$~W\\
        & F & 267{\raisebox{0.5ex}{\tiny$^{+0}_{-1}$}} & $<$0.375{\raisebox{0.5ex}{\tiny$^{+0.001}_{-0}$}}\% & $5.8 \times 10^{13}$~W & & 662{\raisebox{0.5ex}{\tiny$^{+22}_{-19}$}} & $<$0.151{\raisebox{0.5ex}{\tiny$^{+0.004}_{-0.005}$}}\% & $5.7 \times 10^{15}$~W\\
        & G & 337{\raisebox{0.5ex}{\tiny$^{+2}_{-2}$}} & $<$0.297{\raisebox{0.5ex}{\tiny$^{+0.002}_{-0.002}$}}\% & $5.8 \times 10^{13}$~W & & 2677{\raisebox{0.5ex}{\tiny$^{+297}_{-211}$}} & $<$0.037{\raisebox{0.5ex}{\tiny$^{+0.003}_{-0.004}$}}\% & $7.2 \times 10^{15}$~W\\
        & K & 1034{\raisebox{0.5ex}{\tiny$^{+8}_{-10}$}} & $<$0.097{\raisebox{0.5ex}{\tiny$^{+0.001}_{-0.001}$}}\% & $2.4 \times 10^{14}$~W & & 5287{\raisebox{0.5ex}{\tiny$^{+468}_{-415}$}} & $<$0.019{\raisebox{0.5ex}{\tiny$^{+0.002}_{-0.002}$}}\% & $7.0 \times 10^{15}$~W\\
        & M & 66{\raisebox{0.5ex}{\tiny$^{+1}_{-1}$}} & $<$1.52{\raisebox{0.5ex}{\tiny$^{+0.02}_{-0.02}$}}\% & $2.0 \times 10^{14}$~W & & 114{\raisebox{0.5ex}{\tiny$^{+0}_{-0}$}} & $<$0.9{\raisebox{0.5ex}{\tiny$^{+0}_{-0}$}}\% & $8.6 \times 10^{14}$~W\\
        \midrule
        \multirow{4}{*}{Red giant branch} & - & 61{\raisebox{0.5ex}{\tiny$^{+1}_{-0}$}} & $<$1.64{\raisebox{0.5ex}{\tiny$^{+0}_{-0.03}$}}\% & $3.6 \times 10^{13}$~W & & 300{\raisebox{0.5ex}{\tiny$^{+31}_{-16}$}} & $<$0.33{\raisebox{0.5ex}{\tiny$^{+0.02}_{-0.03}$}}\% & $5.7 \times 10^{15}$~W\\
        & G & 4{\raisebox{0.5ex}{\tiny$^{+0}_{-0}$}} & $<$25.0{\raisebox{0.5ex}{\tiny$^{+0}_{-0}$}}\% & $1.4 \times 10^{12}$~W & & 34{\raisebox{0.5ex}{\tiny$^{+4}_{-4}$}} & $<$2.9{\raisebox{0.5ex}{\tiny$^{+0.4}_{-0.3}$}}\% & $5.7 \times 10^{15}$~W\\
        & K & 56{\raisebox{0.5ex}{\tiny$^{+1}_{-0}$}} & $<$1.79{\raisebox{0.5ex}{\tiny$^{+0}_{-0.03}$}}\% & $3.6 \times 10^{13}$~W & & 263{\raisebox{0.5ex}{\tiny$^{+26}_{-12}$}} & $<$0.38{\raisebox{0.5ex}{\tiny$^{+0.02}_{-0.03}$}}\% & $4.6 \times 10^{15}$~W\\
        & M & - & - & - & & 2{\raisebox{0.5ex}{\tiny$^{+0}_{-0}$}} & $<$50.0{\raisebox{0.5ex}{\tiny$^{+0}_{-0}$}}\% & $1.1 \times 10^{15}$~W\\
        \bottomrule
        \label{T3}
    \end{tabular}}
\end{table*}

\section{Conclusions} \label{4}

We have defined an extended sample of stars observed by the recent Breakthrough Listen campaigns (e.g. \citealt{enriquez2017} and \citealt{Price2020}). This has been made possible by using the {\textit{Gaia}} DR2 catalogue to extend the original Breakthrough Listen sample by including stars with measured parallaxes, and inferred distances, that happen to reside within the FWHM of the primary beam response of the GBT and Parkes radio telescopes. This provides a new and much larger stellar sample that increases the number of nearby ($d < 50$~pc) stars by 196{\raisebox{0.5ex}{\tiny$^{+42}_{-7}$}} and extends the total sample to 288315 stars. The extended sample is also significantly enlarged in terms of stellar distances - the median stellar distance of the sample is 2587.8{\raisebox{0.5ex}{\tiny$^{+2090.1}_{-1124.5}$}}~pc.

Adding a large number of stars whose only qualifying criteria are that they lie within the FWHM of the telescope's beam, and have distances inferred from {\textit{Gaia}} parallax measurements, serves to introduce a good mix of stellar systems to the sample, with a much broader range of spectral type. For the first time, we have used this property of the extended sample to place limits on the prevalence of continuous extraterrestrial transmitters on the basis of stellar type.

Adopting the Continuous Waveform Transmitter Rate Figure of Merit (CWTFM) first proposed by \cite{enriquez2017}, and applying this to subsets of our extended sample, leads to figures of merit that are a substantial improvement on any previous survey analysis. More specifically, we conclude that fewer than $\sim$ 0.0660{\raisebox{0.5ex}{\tiny$^{+0.0004}_{-0.0003}$}}\% of the stellar systems within 50~pc possess high duty cycle radio transmitters with EIRP $\gtrsim 10^{13}$~W. At 100 and 200~pc, the incidence is $\lesssim$ 0.061{\raisebox{0.5ex}{\tiny$^{+0.001}_{-0.003}$}}\% (for an EIRP $\gtrsim 6.5 \times 10^{13}$~W), and $\lesssim$ 0.039{\raisebox{0.5ex}{\tiny$^{+0.004}_{-0.008}$}}\% (for an EIRP $\gtrsim 2.5 \times 10^{14}$~W) respectively. These are by far the best limits that have been presented to date for narrow-band signals. 

A major milestone for the Breakthrough Listen Initiative is to survey up to 1 million nearby stars over a continuous frequency range of 580-3500 MHz ($\nu_{\rm rel} \sim 1.4$) with an EIRP$_{\rm min} \sim 2 \times 10^{13}$~W. As shown in Fig. \ref{f12}, the figure of merit of such a survey is significantly better than anything else achieved to date. We note that the million-source survey will need to perform an analysis not unrelated to the approach presented here. In particular, for each MeerKAT pointing, it will be important to identify multiple sources in the large field of view of the individual 13.5-m antennas, selecting targets with measured distances and minimal offsets from the antenna boresight. Since it is expected that MeerKAT will observe other potential (stellar) targets in the field of view by forming multiple phased-array pencil-beams on the sky, the computational expense of this will limit the number of sources that can be observed simultaneously (as compared to a single parabolic telescope). It may be possible to restore the full field-of-view by also adopting wide-field interferometric approaches in SETI surveys \citep{Garrett2018}, alongside beamforming.

Future SETI surveys can generate much better limits on the incidence of artificial radio transmitters of extraterrestrial origin by including in their analysis the effect of observing many other stars in the telescope beam, in addition to any explicit singular target.  In this work we leveraged the {\textit{Gaia}} DR2 catalogue alone, but in future work this approach could be augmented by stellar population synthesis to ensure Milky Way completeness. A further augmentation could add extragalactic population studies to include the more distant stars associated with other galaxies in the field of view. Constraints can also be placed on the prevalence of extraterrestrial transmitters for a wide variety of cosmic objects in the field of view, including various "exotica", from distant AGN to nearby solar system objects (see \citealt{Lacki2020}).  

\section*{Data availability}

This work has made use of data from the European Space Agency (ESA) mission {\textit{Gaia}} (\url{https://www.cosmos.esa.int/gaia}), processed by the {\textit{Gaia}} Data Processing and Analysis Consortium (DPAC, \url{https://www.cosmos.esa.int/web/gaia/dpac/consortium}). The derived data in this article are available in the article and its online supplementary material.

\section*{Acknowledgements}

We gratefully acknowledge the comments of an anonymous reviewer, which were very helpful in preparation of this manuscript. Breakthrough Listen is managed by the Breakthrough Initiatives, sponsored by the Breakthrough Prize Foundation. This work has made use of data from the European Space Agency (ESA) mission
{\textit{Gaia}} (\url{https://www.cosmos.esa.int/gaia}), processed by the {\textit{Gaia}}
Data Processing and Analysis Consortium (DPAC,
\url{https://www.cosmos.esa.int/web/gaia/dpac/consortium}). Funding for the DPAC
has been provided by national institutions, in particular the institutions
participating in the {\textit{Gaia}} Multilateral Agreement. This research has made use of the SIMBAD database,
operated at CDS, Strasbourg, France. This research has made use of "Aladin sky atlas" developed at CDS, Strasbourg Observatory, France. This research has made use of the VizieR catalogue access tool, CDS, Strasbourg, France. This research made use of TOPCAT, an interactive graphical viewer and editor for tabular data \citep{TOPCAT2005}. This research made use of Astropy, a community-developed core Python
package for Astronomy \citep{astropy2013,astropy2018}.

%%%%%%%%%%%%%%%%%%%%%%%%%%%%%%%%%%%%%%%%%%%%%%%%%%



\bibliographystyle{mnras}
\bibliography{ref}

\bsp	% typesetting comment
\label{lastpage}
\end{document}